\documentclass[10pt,conference]{IEEEtran}

\usepackage{cite}
\usepackage{graphicx}
\usepackage{psfrag}
\usepackage{amsmath,amssymb}
\usepackage{algorithmic}
\usepackage{color}

\def\reals{{\mathbb{R}}}
\def\complex{{\mathbb{C}}}
\def\naturals{{\mathbb{N}}}

\def\ints{{\mathbb{Z}}}

\def\bfZero{\mathbf{0}}

\def\bfc{{\mathbf{c}}}

\def\bfs{{\mathbf{s}}}
\def\bfsh{{\hat{\mathbf{s}}}}
\def\bfst{{\tilde{\mathbf{s}}}}
\def\bfsb{{\bar{\mathbf{s}}}}

\def\bfw{{\mathbf{w}}}

\def\bfx{{\mathbf{x}}}

\def\bfy{{\mathbf{y}}}

\def\bfz{{\mathbf{z}}}

\def\bfA{{\mathbf{A}}}

\def\bfF{{\mathbf{F}}}

\def\bfFb{{\bar{\mathbf{F}}}}
\def\bfG{{\mathbf{G}}}

\def\bfH{{\mathbf{H}}}

\def\bfI{{\mathbf{I}}}

\def\bfR{{\mathbf{R}}}

\def\bfRt{{\tilde{\mathbf{R}}}}

\def\bfT{{\mathbf{T}}}

\def\bfW{{\mathbf{W}}}

\def\bfX{{\mathbf{X}}}
\def\bfXh{{\hat{\mathbf{X}}}}

\def\bfY{{\mathbf{Y}}}

\def\bfDelta{{\mathbf{\Delta}}}

\def\Oset{\mathcal{O}}

\def\Sset{\mathcal{S}}

\def\Xset{\mathcal{X}}

\def\hr{^{\mathrm H}}

\def\fro{{\mathrm F}}

\def\vec{\mathrm{vec}}

\newcommand{\prob}[1]{\mathrm{P}\left(#1\right)}
\newcommand{\tprob}[1]{\mathrm{P}(#1)}

\newcommand{\mat}[1]{[{#1}]}

\DeclareMathOperator*{\dotleq}{\overset{.}{\leq}}
\DeclareMathOperator*{\dotgeq}{\overset{.}{\geq}}
\DeclareMathOperator*{\defeq}{\triangleq}

\newtheorem{theorem}{Theorem}
\newtheorem{corollary}[theorem]{Corollary}

\newtheorem{lemma}[theorem]{Lemma}

\def\ml{\mathrm{ML}}
\def\nt{n_\mathrm{T}} % Size
\def\nr{n_\mathrm{R}}

\def\kron{\otimes}

\def\cints{\ints_{\mathcal{G}}}
\def\hr{^\dagger}

\def\nout{\overline{\Oset}_{\epsilon}}
\renewcommand{\mat}{\mathrm{Mat}}

\begin{document}
%\sloppy

\title{LR-aided MMSE lattice decoding is DMT optimal for all approximately universal codes}

\author{\IEEEauthorblockN{Joakim Jald\'{e}n}
\IEEEauthorblockA{Institute of Communications and Radio-Frequency Engineering\\
Vienna University of Technology\\
Vienna, Austria\\
Email: jjalden@nt.tuwien.ac.at}
\and 
\IEEEauthorblockN{Petros Elia}
\IEEEauthorblockA{Departement of Mobile Communications \\ EURECOM\\
Sophia Antipolis, France\\
Email: elia@eurecom.fr}}

\maketitle

\begin{abstract}
Currently for the $\nt \times \nr$ MIMO channel, any explicitly constructed space-time (ST) designs that achieve optimality with respect to the diversity multiplexing tradeoff (DMT) are known to do so only when decoded using maximum likelihood (ML) decoding, which may incur prohibitive decoding complexity. In this paper we prove that MMSE regularized lattice decoding, as well as the computationally efficient lattice reduction (LR) aided MMSE decoder, allows for efficient and DMT optimal decoding of any approximately universal lattice-based code. The result identifies for the first time an explicitly constructed encoder and a computationally efficient decoder that achieve DMT optimality for all multiplexing gains and all channel dimensions. The results hold irrespective of the fading statistics.
\end{abstract}

\vspace{-4pt}
\section{Introduction}

The introduction of MIMO-related scenarios such as MIMO-OFDM and cooperative-diversity has introduced the need for multi-dimensional encoding schemes which can be efficiently decoded and which can guarantee good error probability performance under a plethora of channel topologies and statistics. Towards addressing this need, substantial amounts of research has looked to improve and analyze the error probability performance and decoding complexity of different MIMO encoding/decoding schemes, with such work often focusing on applying specific error probability performance measures to analyze the behavior of specific transmission schemes as they are decoded by different, optimal and suboptimal decoders.

\subsection{Related work} 

From the encoding point of view, substantial research has aimed towards providing space-time (ST) codes with structure that allows for good error probability performance and efficient decoding. Such work include \cite{KR:08} which provides codes based on Clifford algebras that can be seen as generalizations of orthogonal designs and which have good maximum likelihood (ML) decoding complexity. Furthermore, the work in \cite{HR:07} describes codes that take advantage of channel asymmetry ($\nr<\nt$) to achieve good error probability performance with reduced decoding complexity.

From the point of view of detection and decoding, ML-based decoders are known to provide optimal performance but often do so only with prohibitive computational complexity.  Different computationally efficient sub-optimal receiver architectures were introduced, with work focusing on linear receivers (MMSE and ZF) and their decision feedback equalization (DFE) variants \cite{TV:05}, as well as lattice reduction (LR) aided versions of these \cite{YW:02,WBK:04}. Substantial work further looked to analyze the performance of such receivers. For example, the work in \cite{TMK:07} showed that LR-aided ZF decoding can achieve maximal receive diversity for uncoded V-BLAST.

With the recent emergence of the diversity multiplexing tradeoff (DMT) \cite{ZT:03} describing, in a unified manner, the fundamental performance limits of outage-limited MIMO communications, research has focused on establishing the DMT performance of different encoder/decoder architectures.
The work in \cite{TK:07} proved that the naive lattice decoder fails to achieve the diversity multiplexing tradeoff in general.  Furthermore, in \cite{KCM:07}, DMT analysis reveals that both ZF and MMSE linear receivers are suboptimal in terms of their achievable diversity. An important step towards establishing that DMT optimality can be achieved with computationally efficient encoders and decoders was presented in \cite{ECD:04}. By using an ensemble of lattice codes, an MMSE pre-processing step, and an optimal lattice translate, it was shown that there exists lattice codes that, when decoded using lattice decoding, achieve optimal DMT performance over the i.i.d. Rayleigh fading channel.

\subsection{Contributions of present work}
In this work, we extend the results in \cite{ECD:04} by bypassing random ensemble arguments to show that DMT optimality is achievable for all multiplexing gains, by employing explicitly constructed encoders and computationally efficient decoders. Specifically, we consider explicitly constructed approximately universal codes \cite{ESK:07,ORB:06,KC:06}, and regularized lattice decoding.  It is shown that DMT optimality holds for all fading statistics. The key to DMT optimality, as will be shown later, is the MMSE regularization of the decoding metric. We also establish the DMT optimality of the computationally efficient LLL based LR-aided MMSE decoder \cite{WBK:04}.

\section{System model and space-time coding} \label{sec:model}

We consider the quasi-static $\nt \times \nr$ MIMO channel model
\begin{equation} \label{eq:channel}
\bfY = \bfH \bfX + \bfW
\end{equation} 
where $\bfY \in \complex^{\nr \times T}$, $\bfH \in \complex^{\nr \times \nt}$, $\bfX \in \complex^{\nt \times T}$ for $T \geq \nt$, $\bfW \in \complex^{\nr \times T}$, and $\vec(\bfW) \sim \mathcal{N}_{\complex}(\bfZero,\bfI)$. Here, we use $\vec(\bfW)$ to denote the column-by-column vectorization of $\bfW$, and $ \mathcal{N}_{\complex}(\bfZero,\bfI)$ to denote a rotationally invariant circularly symmetric complex normal random vector with unit variance. The code matrices $\bfX$ are drawn from a space-time block code $\Xset$,
satisfying the power constraint
\begin{equation} \label{eq:power-constraint}
\frac{1}{T} \|Ê\bfX \|_{\fro}^2 \leq \rho \, \quad \forall \, \bfX \, \in \Xset \, .
\end{equation}

\subsection{The diversity multiplexing tradeoff}

The rate of an ST code $\Xset$ is given by $R \defeq T^{-1} \log |\Xset|$ and a sequence of codes or \emph{scheme}, indexed by $\rho$, is said to have a \emph{multiplexing gain} of $r$ if (c.f.\ \cite{ZT:03})
$$
r = \lim_{\rho \rightarrow \infty} \frac{R}{\log \rho} \, .
$$
When $\bfXh$ is the output of the decoder (not necessarily ML) given $\bfY$ and $\bfH$, the \emph{diversity gain} $d$ of the scheme is 
$$
d = - \lim_{\rho \rightarrow \infty} \frac{\log \tprob{\bfXh \neq \bfX}}{\log \rho} \, .
$$
The central result of \cite{ZT:03} is that for a fixed multiplexing gain $r$, there is a fundamental limit to the diversity gain: 
\begin{equation} \label{eq:outage}
d \leq d_{\text{out}}(r) \defeq - \lim_{\rho \rightarrow \infty} \frac{\log \tprob{\log \det(\bfI + \rho \bfH\bfH\hr) \leq \bar{R}}}{\log \rho} \, ,
\end{equation}
where $\bar{R} = r \log \rho$ and where $\bfH^\dagger$ denotes the Hermitian transpose of $\bfH$.
In the case of i.i.d.\ Rayleigh fading, $d_{\text{out}}(r)$ is given by the piecewise linear curve connecting  $(\nr-k)(\nt-k)$ for $k=0,1,\ldots,\min(\nr,\nt)$ \cite{ZT:03}. A scheme which satisfies \eqref{eq:outage} with equality for some $r$ is said to be DMT optimal for this multiplexing gain.

We will in the following make use of the $\doteq$ notation (c.f.\ \cite{ZT:03}) for exponential equalities where $f(\rho) \doteq \rho^b$ is taken to mean $\lim_{\rho \rightarrow \infty} \log f(\rho)/\log \rho = b$. The symbols $\dotleq$ and $\dotgeq$ are defined similarly. Let $\nout$ denote the $\epsilon$-no-outage set given by
\begin{equation} \label{eq:no-outage}
\nout \defeq \{ \bfH \, | \, \log \det (\bfI + \rho \bfH\bfH\hr) > (r+\epsilon) \log \rho \} \, .
\end{equation}
As noted in \cite{TV:06} (see also \cite{ZT:03}), a sufficient condition for DMT optimality, regardless of the fading statistics, is that
\begin{equation} \label{eq:sufficient}
\tprob{\bfXh \neq \bfX \, | \, \bfH \in \nout } \doteq \rho^{-\infty}
\end{equation}
for all $\epsilon > 0$, i.e.\ that the conditional probability of decoding error vanishes exponentially fast for channels in $\nout$.

\subsection{Approximately universal lattice space-time codes}

In this paper, we consider a sequence of lattice ST codes: 
\begin{equation} \label{eq:lst-code}
\Xset = \{ \bfX = \mat(\theta \bfG \bfs) \, | \, \bfs \in \Sset_{r} \}
\end{equation}
where $\theta \in \reals_{+}$, $\bfG \in \complex^{\nt T \times \kappa}$ for some
$\kappa \in \naturals$
, and where
\begin{equation} \label{eq:symbol-set}
\Sset_{r} \defeq \{ \bfs \in \cints^\kappa \, | \,  \|Ê\bfs\|^2 \leq \rho^\frac{rT}{\kappa} \} \, ,
\end{equation}
where $\cints = \ints + i \ints$ denotes the set of Gaussian integers\footnote{Extensions of our main results to other constellations, such as the HEX constellations, is straightforward and will appear in a journal version of this work. It is omitted here due to lack of space.}. $\mat(\bfx)$ denotes the $\nt \times T$ matrix formed via column-by-column stacking of consecutive $\nt$-tuples of $\bfx \in \complex^{\nt T}$. Each codeword is thus associated, via the \emph{lattice generator matrix} $\bfG$, to a unique data vector $\bfs \in \Sset_{r} \subset \cints^\kappa$. The choice of $\Sset_{r}$ in \eqref{eq:symbol-set} ensures a multiplexing gain $r$ and choosing $\theta$ in order to satisfy the power constraint (c.f.\ \eqref{eq:power-constraint}) with equality implies that $\theta^2 \doteq \rho^{1-\frac{rT}{\kappa}}$. We assume throughout that the lattice generator matrix $\bfG$ is independent of $\rho$ and $r$.

A key feature of lattice ST codes is that they may be decoded by a class of decoders known as lattice decoders \cite{ECD:04}. To this end we note that the input-output relation from $\bfs$ to $\bfy \defeq \vec(\bfY)$ is
\begin{equation} \label{eq:equivalent-model}
\bfy = \bfF\bfs + \bfw
\end{equation}
where $\bfw \defeq \vec(\bfW)$, and where the \emph{effective channel matrix} is
\begin{equation} \label{eq:equivalent-channel}
\bfF \defeq \theta (\bfI_{T} \kron \bfH)\bfG \, .
\end{equation}
The ML decoder is thus equivalent to (c.f. \cite{ECD:04})
\begin{equation} \label{eq:ml}
\bfsh_{\ml} = \arg \min_{\bfsh \in \Sset_{r}} \| \bfy - \bfF \bfsh \|^2 \, ,
\end{equation}
and may be approximated by a lattice decoder, whereby the constellation boundary imposed by $\Sset_{r}$ is ignored \cite{ECD:04}. The decoding of the lattice ST codes will be discussed in greater detail in Sections \ref{sec:lattice-decoding} and \ref{sec:lr-aided}.

Let $\mu_{i}(\bfA)$ denote the $i$th eigenvalue of a Hermitian matrix $\bfA \in \reals^{q \times q}$, ordered such that $\mu_{1}(\bfA) \leq \ldots \leq \mu_{q}(\bfA)$. Further, let $n \defeq \min(\nt,\nr)$. A sequence of ST codes (not necessarily lattice codes) is \emph{approximately universal} \cite{TV:06} over the $\nt \times \nr$ channel if and only if (c.f.\ \cite{ESK:07})
\begin{equation} \label{eq:approx-universal}
\prod_{i=1}^n \mu_{i}(\bfDelta\bfDelta\hr) \dotgeq \rho^{n-r}
\end{equation}
for all codeword difference matrices $\bfDelta = \bfX_{1}-\bfX_{2}$, where $\bfX_{1},\bfX_{2} \in \Xset$, and $\bfX_{1} \neq \bfX_{2}$. It is known that approximate universality is a sufficient condition for DMT optimality for any fading statistics, assuming ML decoding \cite{TV:06}. For approximately universal codes we have the following lemma that follows directly from \cite[Equation (21)]{TV:06}\footnote{In relation to the result presented here, we point out a small typo in equations (20) and (21) in \cite{TV:06}, where in (20) $2^{R(1+\epsilon)}(|\lambda_{1}|\cdots|\lambda_{n_{m}}| )^{2/n_{m}}$ should be replaced by $(2^{R(1+\epsilon)}|\lambda_{1}|^2\cdots|\lambda_{n_{m}}|^2 )^{1/n_{m}}$, c.f.\ (17) in the same paper. Note also the slightly different definition of $\nout$ in our paper and $\Oset_{\epsilon}$ in \cite{TV:06}, where in the definition of $\nout$ we use $(r+\epsilon)$ in place of $r(1+\epsilon)$.
}.

\vspace{3pt}
\begin{lemma} \label{lm:bound}
Let $\bfDelta = \bfX_{1} - \bfX_{2}$ for $\bfX_{1}, \bfX_{2} \in \Xset$, $\bfX_{1} \neq \bfX_{2}$. If $\Xset$ is approximately universal over the $\nt \times \nr$ channel and $\bfH \in \nout$ it follows that $\| \bfH \bfDelta \|^2_{\fro} \dotgeq \rho^{\frac{\epsilon}{n}}$.
\end{lemma}
\vspace{3pt}

For approximately universal \emph{lattice} ST codes we may also give the following corollary to Lemma \ref{lm:bound}. The proof is given in the appendix.

\vspace{3pt}
\begin{corollary} \label{co:bound}
Let $\Xset$ be a lattice ST code of the form \eqref{eq:lst-code} which, for a fixed lattice generator matrix $\bfG$, is approximately universal for all multiplexing gains in a neighborhood of $r$. Then, for $\bfs_{1},\bfs_{2} \in \Sset_{r+\zeta}$, $\bfs_{1} \neq \bfs_{2}$, and $\bfH \in \nout$ it holds that
\begin{equation} \label{eq:bound}
 \| \bfF (\bfs_{1} - \bfs_{2}) \|^2 \dotgeq \rho^{\frac{\epsilon-\zeta}{n}-\frac{\zeta T}{\kappa}}
\end{equation}
for sufficiently small $\zeta$, $0 < \zeta < \epsilon$.
\end{corollary}
\vspace{3pt}

Approximately universal lattice ST codes, which satisfy the conditions of Corollary \ref{co:bound}, are known to exist for any $(\nr,\nt)$-tuplet and multiplying gain $r$, see e.g.\ \cite{ESK:07}. The codes in \cite{ESK:07} are in fact, for a fixed $\bfG$, approximately universal over all $r \in [0,n]$. In what follows we only consider codes for which Corollary \ref{co:bound} applies. We also point out that in the definition of approximately universal lattice ST codes we require that the set of data symbols $\Sset_{r}$ is given by the Gaussian integers within a hyper-sphere of radius $\rho^{\frac{rT}{2\kappa}}$. It is readily seen that presented analysis carries over  (at the expense of extra notational complexity) to the more practical case where the constellation is cubic, i.e.\ $|\Re(s_{k})|, |\Im(s_{k})| \leq \rho^{\frac{rT}{2\kappa}}$,
which also maintains the scheme's multiplexing gain.

\section{Regularized lattice decoding} \label{sec:lattice-decoding}

As noted in Section \ref{sec:model}, ML decoding is equivalent to solving \eqref{eq:ml}. The \emph{naive} lattice decoder (c.f.\ \cite{ECD:04}) is obtained by simply ignoring the constellation boundary of $\Sset_{r} \subset \cints^\kappa$:
\begin{equation} \label{eq:naive}
\bfsh_{0} = \arg \min_{\bfsh \in \cints^\kappa} \| \bfy - \bfF \bfsh \|^2 \, .
\end{equation}
We count the event when the decoder decides in favor of a codeword not in the constellation as an error. The benefit of using \eqref{eq:naive} in place of \eqref{eq:ml} is that one may avoid the potentially complicated boundary control, and apply tools from lattice reduction theory for solving \eqref{eq:naive}. However, as argued in \cite{ECD:04} and subsequently proved in \cite{TK:07}, the naive lattice decoder is not in general DMT optimal. It was however also shown in \cite{ECD:04} that the problem is not with lattice coding and decoding per se, but rather with the naive implementation.

Intuitively, as the ML decoder (c.f.\ \eqref{eq:ml}) is DMT optimal for approximately universal codes and the naive lattice decoder is not, the sub-optimality of the naive lattice decoder must stem from the fact that it decides, with high probability, in favor of codewords that do not belong to the constellation $\Sset_{r}$. Note here that $\bfs \notin \Sset_{r}$, $\bfs \in \cints^\kappa$, implies $\| \bfs \|^2 > \rho^{\frac{rT}{\kappa}}$. Thus, having the decoder penalize vectors $\bfs$ with large norm, one can expect to reduce the probability of out-of-constellation errors. This amounts to regularization of the decoding metric and we let the $\alpha$-regularized lattice decoder be given by
\begin{equation} \label{eq:regularized}
\bfsh_{\alpha} = \arg \min_{\bfsh \in \cints^\kappa} \| \bfy - \bfF \bfsh \|^2 + \alpha \| \bfs \|^2 \, .
\end{equation}
Clearly, for $\alpha = 0$ the regularized lattice decoder coincides with the naive lattice decoder. We will however in what follows show that by choosing $\alpha$ appropriately, one can achieve DMT optimality for any approximately universal code. The result is captured by the following theorem.

\vspace{3pt}
\begin{theorem} \label{thrm:main}
Approximately universal lattice codes, decoded using the $\alpha$-regularized decoder with $\alpha = \rho^{-\frac{rT}{\kappa}}$, achieve DMT optimality and do so irrespective of the fading statistics.
\end{theorem}
\vspace{3pt}

\noindent \emph{Proof:} We will show that when $\bfs$ is the data vector corresponding to the transmitted codeword of an approximately universal code, when $\alpha = \rho^{-\frac{rT}{\kappa}}$ and when $\epsilon > 0$, using the $\alpha$-regularized decoder in \eqref{eq:regularized} implies that $\prob{\bfsh_{\alpha} \neq \bfs | \bfH \in \nout} \doteq \rho^{-\infty}$. In other words, the conditional probability of error vanishes exponentially fast for channels (strictly) not in outage, establishing DMT optimality.

Towards this end, given $\epsilon > 0$, choose $\zeta$ and $\delta$ such that $0 < \zeta < \epsilon$, where $\zeta$ is sufficiently small for Corollary \ref{co:bound} to apply, and such that (c.f.\ \eqref{eq:bound})
\begin{equation} \label{eq:zd-bound}
\frac{\epsilon-\zeta}{n}-\frac{\zeta T}{\kappa} > \delta > 0
\quad \text{and} \quad
\frac{\zeta T}{\kappa} > \delta > 0 \, .
\end{equation}
This can always be done. Assume also that $\bfH \in \nout$ and that the noise vector $\bfw$ satisfies $\| \bfw \|^2 \leq \rho^{\delta}$.

Consider first the $\alpha$-regularized metric (c.f.\ \eqref{eq:regularized}) for the transmitted data vector $\bfs \in \Sset_{r}$. As (c.f.\ \eqref{eq:equivalent-model})
$$
\| \bfy - \bfF \bfs \|^2Ê= \| \bfw \|^2
$$
it follows that
\begin{equation} \label{eq:t-bound}
\| \bfy - \bfF \bfs \|^2Ê+ \alpha \| \bfs \|^2 \leq \rho^{\delta} + \alpha \rho^{\frac{rT}{\kappa}} \doteq \rho^{\delta}
\end{equation}
where we used that $\| \bfw \|^2 \leq \rho^{\delta}$, that $\alpha = \rho^{-\frac{rT}{\kappa}}$ and that $\bfs \in \Sset_{r}$ which implies $\| \bfs \|^2 \leq \rho^{\frac{rT}{\kappa}}$ (c.f.\ \eqref{eq:symbol-set}).

For any data vector $\bfsh \in \Sset_{r+\zeta}$, $\bfsh \neq \bfs$, we note that
$$
\| \bfy - \bfF \bfsh \| = \| \bfF(\bfs-\bfsh) + \bfw \| \geq \| \bfF(\bfs-\bfsh) \| - \| \bfw \| \, .
$$
As $ \| \bfF(\bfs-\bfsh) \| \dotgeq \rho^{\frac{1}{2}(\frac{\epsilon-\zeta}{n}-\frac{\zeta T}{\kappa})}$ by Corallary \ref{co:bound} and as $\| \bfw \| \leq \rho^{\frac{1}{2}\delta}$, it follows by \eqref{eq:zd-bound} that
$$
\| \bfy - \bfF \bfsh \|^2 \dotgeq \rho^{\frac{\epsilon-\zeta}{n}-\frac{\zeta T}{\kappa}}
$$
and
\begin{equation} \label{eq:o-bound-1}
\| \bfy - \bfF \bfsh \|^2 + \alpha \|Ê\bfsh \|^2 \dotgeq \rho^{\frac{\epsilon-\zeta}{n}-\frac{\zeta T}{\kappa}} \, ,
\end{equation}
for any $\bfsh \in \Sset_{r+\zeta}$, $\bfsh \neq \bfs$.

For $\bfsh \notin \Sset_{r+\zeta}$, $\bfsh \in \cints$, it holds that $\| \bfsh \|^2 > \rho^{\frac{(r+\zeta)T}{\kappa}}$ (c.f.\ \eqref{eq:symbol-set}) by which it follows that
\begin{equation} \label{eq:o-bound-2}
\| \bfy - \bfF \bfsh \|^2 + \alpha \|Ê\bfsh \|^2 \geq \alpha \rho^{\frac{(r+\zeta)T}{\kappa}}Ê\geq \rho^{\frac{\zeta T}{\kappa}} \, .
\end{equation}
By defining
$$
\xi \defeq \min\Big( \frac{\epsilon-\zeta}{n}-\frac{\zeta T}{\kappa} \, , \, \frac{\zeta T}{\kappa} \Big)
$$
where $\xi > \delta$ due to \eqref{eq:zd-bound}, and combining \eqref{eq:o-bound-1} and \eqref{eq:o-bound-2} it follows that
\begin{equation} \label{eq:o-bound}
\| \bfy - \bfF \bfsh \|^2 + \alpha \|Ê\bfsh \|^2 \dotgeq \rho^{\xi}
\end{equation}
for any $\bfsh \in \cints^\kappa$, $\bfsh \neq \bfs$. As $\delta < \xi$ (i.e.\ $\delta$ is strictly smaller than $\xi$), it follows by \eqref{eq:t-bound} and \eqref{eq:o-bound} that there is $\rho_{0}$ such that
$$
\| \bfy - \bfF\bfs \|^2 + \alpha \|Ê\bfs \|^2 < \| \bfy - \bfF\bfsh \|^2 + \alpha \| \bfsh \|^2
$$
for any $\bfsh \in \cints^\kappa$, $\bfsh \neq \bfs$, and $\rho \geq \rho_{0}$. This implies that the $\alpha$-regularized decoder will make a correct decision. In other words, if $\bfH \in \nout$, it follows that $\| \bfw \|^2 > \rho^{\delta}$ constitutes a necessary condition for an error to occur when $\rho \geq \rho_{0}$. However, as $\prob{\| \bfw \|^2 \geq \rho^{\delta}} \doteq \rho^{-\infty}$ due to the exponential tails of the Gaussian distribution, we see that $\prob{\bfsh_{\alpha} \neq \bfs | \bfH \in \nout} \doteq \rho^{-\infty}$ and the claim of Theorem \ref{thrm:main} follows. \hfill $\square$

The metric in \eqref{eq:regularized} is not identical to the metric used in the MMSE-GDFE decoder considered in \cite{ECD:04}, although the two metrics share some key features. In particular, if the lattice translate is omitted, it can be shown that the metric in \cite{ECD:04} is equivalent\footnote{Note also that the metric in \cite{ECD:04} is expressed in a real valued form which allows for more general code designs. The real valued reformulation will be considered in a journal version of this work.} to (c.f.\ \eqref{eq:regularized})
\begin{equation} \label{eq:mmse-gdfe}
\| \bfy - \bfF\bfs \|^2 + \rho^{-1} \|Ê\theta \bfG \bfs \|^2 \, ,
\end{equation}
i.e.\ the regularization is applied to the vectorized codeword $\bfx = \theta \bfG \bfs$ instead of $\bfs$. It is a straightforward exercise to repeat the proof of Theorem \ref{thrm:main} and show that decoding with respect to \eqref{eq:mmse-gdfe} is also DMT optimal. To this end, note that $\theta^2\rho^{-1}\doteq \rho^{-\frac{rT}{\kappa}}$. In fact, when $\bfG$ is an orthogonal matrix, as is the case for perfect codes \cite{ORB:06}, \eqref{eq:mmse-gdfe} reduces to \eqref{eq:regularized}. This confirms the observation made in \cite{ECD:04} that the ``magic'' ingredient of the GDFE-MMSE decoder, in terms of DMT optimality, is MMSE pre-processing. Similarly, it reveals that $\alpha = \rho^{-\frac{rT}{\kappa}}$ is the corresponding ``magic'' parameter for the regularized lattice decoder which motivates us to refer to the regularized lattice decoder with $\alpha = \rho^{-\frac{rT}{\kappa}}$ as the MMSE regularized lattice decoder. It should however be noted that the choice of $\alpha = \rho^{-\frac{rT}{\kappa}}$ can naturally also be directly obtained from the linear MMSE filter for $\bfs$ given the observation $\bfy$ (c.f.\ \eqref{eq:equivalent-model} and Section \ref{sec:lr-aided}).

\section{Lattice reduction aided decoding} \label{sec:lr-aided}

By ``completing the squares'', the $\alpha$-regularized metric may equivalently be written as
\begin{equation} \label{eq:squares}
\| \bfy - \bfF \bfsh \|^2 + \alpha \|Ê\bfsh \|^2 = \|\bfz - \bfR \bfsh \|^2 + c
\end{equation}
where $\bfR \in \complex^{\kappa \times \kappa}$ is a square root factor of $\bfF\hr\bfF + \alpha \bfI$, i.e.
\begin{equation} \label{eq:sqrf}
\bfR\hr \bfR = \bfF\hr\bfF + \alpha \bfI \, ,
\end{equation}
where $\bfz \defeq \bfR^{-\dagger}\bfF\hr\bfy$, and where 
\begin{equation}
c \defeq \bfy\hr \big[ \bfI - \bfF\hr(\bfF\hr\bfF+\alpha \bfI)^{-1} \bfF \big] \bfy \geq 0 \, .
\end{equation}
The $\alpha$-regularized decoder can thus be expressed as
\begin{equation} \label{eq:regularized-2}
\bfsh_{\alpha} = \arg \min_{\bfsh \in \cints^\kappa} \| \bfz - \bfR \bfsh \|^2 \, .
\end{equation}
The optimization problem in \eqref{eq:regularized-2} however still require the solution to a closest vector problem (CVP), which is NP-hard in general. This makes sub-optimal solutions appealing. To this end, consider the decoder given by
\begin{equation} \label{eq:mmse}
\bfsh_{\alpha,\mathrm{MMSE}} = \arg \min_{\bfsh \in \cints^\kappa} \| \bfR^{-1}\bfz - \bfsh \|^2 \, .
\end{equation}
The decoder in \eqref{eq:mmse} is easily implemented by component-wise rounding of $\bfR^{-1} \bfz$ to the nearest integer vector. It is relatively straightforward to verify that
$$
\bfR^{-1} \bfz = (\bfF\hr \bfF + \alpha \bfI)^{-1} \bfF\hr \bfy
$$
which implies that the solution to \eqref{eq:mmse} corresponds to the standard linear MMSE decoder.

Yao and Wornell \cite{YW:02} suggested the use of lattice reduction to improve the approximation quality when replacing \eqref{eq:regularized-2} by \eqref{eq:mmse}. The key idea behind this approach is to note that \eqref{eq:regularized-2} is equivalent to
\begin{equation} \label{eq:lr}
\min_{\bfst \in \cints^\kappa} \| \bfz - \bfR \bfT \bfst \|^2
\end{equation}
where $\bfT$ is a unimodular matrix, i.e.\ $\bfT$ is a one-to-one map from $\cints^\kappa$ to $\cints^\kappa$ or equivalently, $\bfT \in \cints^{\kappa \times \kappa}$ and $|\det(\bfT)| = 1$. We write $\bfRt = \bfR \bfT$ in what follows, and refer to $\bfRt$ as the lattice reduced channel. The process of finding $\bfT$, given $\bfR$, is known as lattice reduction.

The sub-optimal solution corresponding to \eqref{eq:lr} is given by
\begin{equation} \label{eq:lr-mmse}
\bfst_{\alpha,\mathrm{LR-MMSE}} = \arg \min_{\bfst \in \cints^\kappa} \| \bfRt^{-1}\bfz - \bfst \|^2 \, 
\end{equation}
where $\bfRt = \bfR \bfT$ and the approximate solution to \eqref{eq:regularized-2} is
\begin{equation} \label{eq:basis-change}
\bfsh_{\alpha,\mathrm{LR-MMSE}} = \bfT \bfst_{\alpha,\mathrm{LR-MMSE}} \, .
\end{equation}
The key observation of \cite{YW:02} is that by making $\bfRt$ well conditioned (by the appropriate choice of $\bfT$), the quality of the approximation may be significantly improved. The resulting decoder (defined by \eqref{eq:lr-mmse} and \eqref{eq:basis-change}) is known as the LR-aided MMSE decoder\cite{WBK:04}.

The most commonly considered lattice reduction algorithm is the computationally efficient LLL algorithm \cite{LLL:82}. The LLL algorithm is also known to provide maximum receive diversity, at multiplexing gain $r=0$ and under i.i.d.\ Rayleigh fading, for uncoded V-BLAST transmissions \cite{TMK:07}. In what follows we prove that LLL based LR-aided decoding can in fact achieve the most general diversity-related optimality, by showing that the LLL based LR-aided MMSE decoder can, in the context of lattice codes, achieve the maximal diversity gain for all multiplexing gains $r$ and fading statistics.

\vspace{3pt}
\begin{theorem}
Approximately universal lattice codes, when decoded using the LLL based LR-aided MMSE decoder, achieve the optimal DMT tradeoff, and do so irrespective of fading statistics.
\end{theorem}
\vspace{3pt}

\noindent \emph{Proof:} To prove the above, we will demonstrate that $\prob{\bfsh_{\alpha,\mathrm{LR-MMSE}} \neq \bfs \, | \, \bfH \in \nout} \doteq \rho^{-\infty}$. To this end, let $\bfRt = \bfR \bfT$ be the LLL lattice reduced channel matrix. It follows by the bounded orthogonality defect of LLL reduced bases (c.f.\ \cite{LLL:82} and the proof in \cite{TMK:07}) that there is a constant $K_{\kappa} > 0$, independent of $\bfR$, for which
\begin{equation} \label{eq:l-bound}
\sigma_{\max}(\bfRt^{-1}) \leq \frac{K_{\kappa}}{\lambda(\bfR)} \, .
\end{equation}
where $\sigma_{\max}(\bfRt^{-1})$ is the largest singular value of $\bfRt^{-1}$ and where
\begin{equation} \label{eq:s-vec}
\lambda(\bfR) \defeq \min_{\bfc \in \cints^\kappa \backslash \{ \bfZero \}} \| \bfR \bfc \|
\end{equation}
denotes the shortest vector in the lattice generated by $\bfR$. Although the proof in \cite{TMK:07} was given for real valued bases it straightforwardly extends to the complex case, c.f.\ \cite{GM:05}.

Assume, as in the proof of Theorem \ref{thrm:main}, that $\bfH \in \nout$ and $\| \bfw \|^2 \leq \rho^{\delta}$. For $\bfsh \in \cints^\kappa$, $\bfsh \neq \bfs$, it follows that
\begin{align*}
\| \bfz - \bfR \bfsh \| = & \; \|Ê(\bfz-\bfR\bfs) + \bfR(\bfs - \bfsh) \| \\
\leq & \; \|Ê\bfR(\bfs - \bfsh) \| + \| \bfz-\bfR\bfs \|
\end{align*}
and
\begin{align}
\| \bfR (\bfs - \bfsh) \| \geq & \; \| \bfz - \bfR \bfsh \| - \| \bfz - \bfR \bfs \| \nonumber \\
\dotgeq & \; (\rho^{\xi} - c)^\frac{1}{2} - \| \bfz - \bfR \bfs \| \label{eq:tmp1}
\end{align}
where the last inequality follows by combining \eqref{eq:o-bound} and \eqref{eq:squares}. As $c \dotleq \rho^{\delta}$ and $\| \bfz - \bfR \bfs \|^2 \dotleq \rho^{\delta}$ by \eqref{eq:t-bound} and \eqref{eq:squares}, and since $\xi > \delta$, we may conclude from \eqref{eq:tmp1} that $\| \bfR(\bfs - \bfsh) \|^2 \dotgeq \rho^{\xi}$, for any $\bfsh \in \cints^\kappa$, $\bfsh \neq \bfs$. By identifying $\bfc = \bfs-\bfsh \in \cints \backslash \{Ê\bfZero \}$ in \eqref{eq:s-vec} it follows that $\lambda^2(\bfR) \dotgeq \rho^{\xi}$ and by \eqref{eq:l-bound} that
\begin{equation} \label{eq:sigma-bound}
\sigma_{\max}^{2}(\bfRt^{-1}) \dotleq \rho^{-\xi} \, .
\end{equation}

From \eqref{eq:lr-mmse} and \eqref{eq:basis-change} it may be seen that $\bfsh_{\alpha,\mathrm{LR-MMSE}} \neq \bfs$ if and only if $\bfst_{\alpha,\mathrm{LR-MMSE}} \neq \bfsb$ where $\bfsb = \bfT^{-1} \bfs$. The metric in \eqref{eq:lr-mmse}, evaluated for $\bfst = \bfsb$, satisfies
\begin{align}
\|Ê\bfRt^{-1} \bfz - \bfsb \|^2 = &  \; \| \bfRt^{-1}(\bfz - \bfRt \bfsb) \|^2 \nonumber \\
\leq & \; \sigma_{\max}^2(\bfRt^{-1}) \|Ê\bfz - \bfRt \bfsb \|^2 \dotleq \rho^{\delta-\xi} \label{eq:lr-bound}
\end{align}
where the last inequality follows by \eqref{eq:sigma-bound} together with $\| \bfz - \bfR \bfs \|^2 \dotleq \rho^{\delta}$ and $\bfR \bfs = \bfRt \bfsb$. For $\bfst \in \cints^\kappa$, $\bfst \neq \bfsb$, it follows that
\begin{align*}
\|Ê\bfRt^{-1} \bfz - \bfst \| = & \; \| \bfRt^{-1} \bfz - \bfsb + (\bfsb-\bfst) \| \\
\geq & \; \| \bfsb - \bfst \| - \| \bfRt^{-1} \bfz - \bfsb \|
\end{align*}
By noting that $\| \bfsb - \bfst \|^2 \geq 1$ if $\bfsb \neq \bfst$, that $\|\bfRt^{-1} \bfz - \bfsb \|^2 \dotleq \rho^{\delta - \xi}$ (c.f.\ \eqref{eq:lr-bound}) and that $\delta - \xi < 0$, it follows that
\begin{equation} \label{eq:lr-bound2}
\|Ê\bfRt^{-1} \bfz - \bfst \|^2 \dotgeq \rho^{0}
\end{equation}
for $\bfst \in \cints^\kappa$, $\bfst \neq \bfsb$. Combining \eqref{eq:lr-bound} and \eqref{eq:lr-bound2} yields
$
\|Ê\bfRt^{-1} \bfz - \bfsb \|^2 < \|Ê\bfRt^{-1} \bfz - \bfst \|^2
$
for all $\bfst \in \cints^\kappa$, $\bfst \neq \bfsb$, and sufficiently large $\rho$ implying that the decision of the LR-aided MMSE decoder (c.f.\ \eqref{eq:lr-mmse} and \eqref{eq:basis-change}) is correct. As in the proof of Theorem \ref{thrm:main}, we see that given $\bfH \in \nout$ it must hold that $\| \bfw \|^2 > \rho^{\delta}$ for an error to occur, which implies
$\prob{\bfsh_{\alpha,\mathrm{LR-MMSE}} \neq \bfs \, | \, \bfH \in \nout} \doteq \rho^{-\infty}$. \hfill $\square$

\section{Conclusion}

In this paper, we consider the problem of efficiently decoding approximately universal lattice ST codes. We show that MMSE regularized lattice decoding in general, and the computationally efficient LLL based LR-aided MMSE decoder in particular, realize the maximum receive diversity and thus DMT optimality for approximately universal lattice codes. The result holds for any fading statistics and confirms that the key to achieving DMT optimality is the regularization of the decoding metric provided by the MMSE decoder.

\appendix

\noindent \emph{Proof of Corollary \ref{co:bound}:} By the equivalent channel model (c.f.\ \eqref{eq:lst-code}, \eqref{eq:equivalent-model} and \eqref{eq:equivalent-channel}) and Lemma \ref{lm:bound} it follows that
$$
\| \bfH (\bfX_{1}-\bfX_{2})\|^2_{\fro} = \| \bfF (\bfs_{1} - \bfs_{2}) \|^2 \dotgeq \rho^{\frac{\epsilon}{n}}
$$
for $\bfs_{1}, \bfs_{2} \in \Sset_{r}$, $\bfs_{1} \neq \bfs_{2}$, given that $\bfH \in \nout$. For the \emph{un-normalized} equivalent channel $\bfFb \defeq (\bfI_{T} \kron \bfH) \bfG$ we have
$$
\theta^2  \| \bfFb (\bfs_{1} - \bfs_{2}) \|^2 = 
\rho^{1-\frac{rT}{\kappa}} \| \bfFb (\bfs_{1} - \bfs_{2}) \|^2 \dotgeq \rho^{\frac{\epsilon}{n}} \, ,
$$
where $\bfFb$ is independent of $\rho$ and $r$ (note that $\bfF = \theta \bfFb$). Consider now the application of Lemma \ref{lm:bound} to a scheme with multiplexing gain $r' = r+\zeta$, where $0 < \zeta < \epsilon$. By the assumption that $\bfH \in \nout$ it follows that
$$
\log \det (\bfI + \rho \bfH\bfH\hr) > (r+\epsilon) \log \rho = (r'+\epsilon-\zeta) \log \rho
$$
which by the application of Lemma \ref{lm:bound} implies that
\begin{equation} \label{eq:tmp-scale}
\rho^{1-\frac{r' T}{\kappa}} \| \bfFb (\bfs_{1} - \bfs_{2}) \|^2 \dotgeq \rho^{\frac{\epsilon-\zeta}{n}}
\end{equation}
for $\bfs_{1},\bfs_{2} \in \Sset_{r'}$, $\bfs_{1} \neq \bfs_{2}$. Rewriting \eqref{eq:tmp-scale} it terms of $r$ yields
$
\rho^{1-\frac{r T}{\kappa}} \| \bfFb (\bfs_{1} - \bfs_{2}) \|^2 \dotgeq \rho^{\frac{\epsilon-\zeta}{n}-\frac{\zeta T}{\kappa}}
$
or equivalently
$
\| \bfF (\bfs_{1} - \bfs_{2}) \|^2 \dotgeq \rho^{\frac{\epsilon-\zeta}{n}-\frac{\zeta T}{\kappa}}
$
for any $\bfs_{1},\bfs_{2} \in \Sset_{r'}= \Sset_{r+\zeta}$, $\bfs_{1} \neq \bfs_{2}$. \hfill $\square$

\vspace{-4.5pt}
\section*{Acknowledgment}
This work was supported by the European Commission through the FP6 STREP project MASCOT (IST-026905) and in the framework of the FP7 Network of Excellence in Wireless COMmunications NEWCOM++ (IST-216715).

\vspace{-0.1pt}

% References

\end{document}